# Room temperature ferromagnetic-like behavior in Mn-implanted and post-annealed InAs layers deposited by Molecular Beam Epitaxy

R. González-Arrabal[a], Y. González[a], L. González[a], M. García-Hernández[b], F. Munnik[c] and M. S. Martín-González[a*]

[a] Instituto de Microelectrónica de Madrid CSIC C/ Isaac Newton, 8. Tres Cantos, E-28760-Madrid, Spain

[b] Instituto de Ciencia de Materiales de Madrid (CSIC), Cantoblanco, E-28049 Madrid, Spain

[c] Institute of Ion Beam Physics and Materials Research, Forschungszentrum Dresden−Rossendorf, P.O. Box 510119, D-01314 Dresden, Germany

* Corresponding author: marisol@imm.cmm.csic.es

We report on the magnetic and structural properties of Ar and Mn implanted InAs epitaxial films grown on GaAs (100) by Molecular Beam Epitaxy (MBE) and the effect of Rapid Thermal Annealing (RTA) for 30 seconds at 750ºC. Channeling Particle Induced X- ray Emission (PIXE) experiments reveal that after Mn implantation almost all Mn atoms are subsbtitutional in the In-site of the InAs lattice, like in a diluted magnetic semiconductor (DMS). All of these samples show diamagnetic behavior. But, after RTA treatment the Mn-InAs films exhibit room-temperature magnetism. According to PIXE measurements the Mn atoms are no longer substitutional. When the same set of experiments were performed with As as implantation ion all of the layers present diamagnetism without exception. This indicates that the appearance of room-temperature ferromagnetic-like behavior in the Mn-InAs-RTA layer is not related to lattice disorder produce during implantation, but to a Mn-reaction produced after a short thermal treatment. X-ray diffraction patterns (XRD) and Rutherford Back Scattering (RBS) measurements evidence the segregation of an oxygen deficient-$MnO_2$ phase (nominally $MnO_{1.94}$) in the Mn-InAs-RTA epitaxial layers which might be on the origin of room temperature ferromagnetic-like response observed.

# INTRODUCTION

Combining semiconducting and ferromagnetic-like properties into the same material provides the opportunity of controlling conduction not only in the traditional way (electronics), but controlling its quantum spin state. Cooperative effects through spin-exchange interaction such as the tuning of the magnetic properties of the semiconductors by changing their carrier density[1] open the way towards to the emerging technology known as magnetoelectronics or spintronics.

At this point, diluted magnetic semiconductors (DMS) have recently been a hot topic in the magnetic semiconductor research. They are based on traditional semiconductors, but are doped with transition metals like Mn. One of the drawbacks of DMS for industrial applications is its commonly low Curie temperature ($T_c$). However, theoretical predictions[2] indicate that $T_c$ can be raised above 300 K for those semiconductors that contain a high concentration of magnetic ions or consist of light elements, like in the case of Mn or Co doped-ZnO. Meanwhile, extensive doubts persist on the origin of their magnetism[3,4,5,6] or on the effective interaction between magnetism and semiconductivity. The III-V semiconductors are the set of materials in which spin injection into non-magnetic semiconductor and ferromagnetism control by light and electric fields for low Mn concentrations, has been demonstrated. Moreover, this kind of materials has also been reported to posses $T_c$ above room temperature, i.e. for $In_{1-x}Mn_xAs$ samples grown by metal-organic vapor phase epitaxy a $T_c \sim 330$ K[7, 8] has been reported. Although, in some other works an experimental $T_c < 55$ K has been measured for this phase.[9] Incorporation of a large amount of magnetic atoms in a controlled way during the growth of III-V semiconductor structures by MBE is a difficult task. Usually, the low equilibrium solubility of transition metals (such as Mn) into III-V semiconductors, originates surface segregation or phase separation above a certain doping threshold. In the case of InAs, magnetic dopants were successfully introduced during growth for the first time in 1989 by Ohno et al. who deposited $In_{1-x}Mn_xAs$ layers by MBE at relatively low temperatures (200-300ºC) and Mn concentrations ($x \leq 0.026$).[10] To overcome these difficulties, implantation techniques that allow insertion of magnetic impurities into the semiconductors in a controlled way, were started to be used.[11,12,13]

Even when the $In_{1-x}Mn_xAs$ system has been relatively well characterized, some controversies are still found in literature on the origin of the magnetic properties. $In_{1-x}Mn_xAs$ layers were found to be ferromagnetic or paramagnetic depending on the local environment surrounding the Mn atoms, which in turn depends on the growth procedure.[10,14,15] For films grown at relatively low temperature (~200ºC) Mn has been observed to incorporate into the InAs structure occupying the In-site forming an homogeneous InMnAs alloy which presents a paramagnetic behavior, while for films deposited at temperatures higher than 300ºC Mn might form MnAs clusters which are supposed to be the responsible of the ferromagnetic behavior of these films.[15]

To elucidate this point we examine the origin of the magnetism in Mn-implanted InAs epitaxial grown by MBE and the effect of rapid thermal annealing. The structure of the InAs crystal lattice is investigated for the different processes by X-ray Diffraction. The layer composition has been characterized by Rutherford Backscattering Spectroscopy (RBS). The relation of magnetism with lattice location of Mn within the crystal structure is investigated by particle induced X-ray emission (PIXE) and scanning quantum interference device (SQUID) measurements. The appearance of secondary phases in the structure after annealing and its relation with the magnetic properties of the samples is investigated and discussed.

EXPERIMENTAL

InAs layers with a thickness of 250 nm were grown on commercial GaAs (100) substrates by ALMBE[16] at a substrate temperature Ts = 440ºC, a growth rate rg = 0.5 monolayers per second (ML/s) and an arsenic beam equivalent pressure BEP (As4) = $2x10^6$ Torr. Previous to the growth of InAs, a 200 nm thick GaAs buffer layer was grown at $T_s$= 580ºC and rg=1ML/s by MBE. According to the reflection high energy electron (RHEED) pattern, a two dimensional InAs growth front is achieved after the first 30 nm were grown. InAs samples were implanted with Mn ions at fluences of $1x10^{16}$ $cm^{-2}$ and $8x10^{16}$ $cm^{-2}$. InAs samples implanted with Ar ions at a fluence of $1x10^{16}$ $cm^{-2}$ have also been studied for comparison purposes. In order to avoid implantation-induced sputtering of the InAs layer, the Mn implantation at high dose ($8x10^{16}$ $cm^{-2}$) was sequentially carried out in four steps, each of them at a fluence of $2x10^{16}$ $cm^{-2}$. The implantations were performed at room temperature and out of channeling incidence. The implantation conditions were selected to achieve a similar damage pattern by implanting with Mn and with Ar. The implantation energies were 190 KeV and 180 KeV for Mn and for Ar, respectively. Samples were rapid thermal annealed in Ar atmosphere for 30 seconds at a temperature of 750 ºC. During annealing, the sample surface was covered with an InAs/GaAs sample to prevent arsenic evaporation.

Structural characterization of the samples was carried out by X-Ray Diffraction using a Philips X-PERT four cycle diffractometer with a Cu $K_\alpha$ radiation source. The XRD measurements were performed in Bragg-Brentano geometry.

The layer composition has been characterized by Rutherford Backscattering Spectroscopy (RBS). RBS measurements were performed using very small currents of ~2nA of a 2 MeV $He^+$ beam and placing a standard Si surface barrier detector at 170º with respect to the beam direction.

Proton Induced X-ray (PIXE) measurements in channeling configuration (PIXE/C) were carried out to study the lattice location of the Mn atoms within the InAs host matrix. PIXE angular scan curves were measured using an $H^+$ beam at energy of 1.5 MeV. The beam current for PIXE measurements was close to 1 nA and the spot size was about 1.0 x1.2 $mm^2$. The total beam fluence for a single analysis scan was around 140nC to avoid radiation-induced damage during the measurements. Characteristic X-rays were measured by a Si(Li) detector located at an angle of 135º with respect to the beam direction. The position of the Mn atoms in the host matrix has been determined by comparing the shape and position of the Mn angular scan curve with that for In. The lattice disorder and the percentage of host atoms displaced after implantation, as well as the damage recovery after RTA have been characterized by the minimum yield ($\chi_{min}$) from the measured PIXE angular scan curves. The minimum yield is the ratio of the yield measured in channeling direction to that measured in random configuration. $\chi_{min}$ scales with lattice disorder being a good indicator of it.

Magnetic characterization of the samples was performed by scanning quantum interference device (SQUID). The temperature dependence of the magnetization at constant field and the field dependence at constant temperature were measured in the cero field cold (ZFC) protocol. The field dependence of the magnetization was measured at temperatures of 5 and 300K.

RESULTS AND DISCUSSION

The projected ranges of the implanted Mn and Ar ions are calculated with the TRIM code[17] to be approximately the same for Mn and Ar. As shown in Fig.1 the elemental depth distribution of these elements within the InAs film is almost identical presenting a Gaussian-like shape with the center of the Gaussian located at the center of the film (125nm). The maximum number of target atom displacements per ion is calculated to be 3.7 and 2.2 for Mn- and Ar-implanted layers, respectively.

The temperature dependence of the magnetization at constant field for the as-grown (reference), Mn-implanted and Mn-implanted+RTA (from now on Mn-RTA) as well as, for Ar-implanted and Ar-RTA samples is shown in Fig. 2. The reference film exhibits a diamagnetic behavior with a magnetic susceptibility of around $-3x10^{-6}$ emu/gOe. Similar behavior is observed for Mn-implanted, Ar-implanted and Ar-RTA samples. Notice that the diamagnetic response seems to decrease when samples are implanted, regarding the reference sample. We relate the observed behavior to the implantation induced damage as evidenced in the broadening of the (200) InAs peak (see Fig. 4) and because a similar effect is observed for the Ar-implanted samples. The Mn-RTA sample is the only one presenting positive magnetic susceptibility values for the whole temperature range studied. The fact that at 300 K the magnetization for the Mn-RTA sample drops to about 68% of its low temperature value would imply that the Curie temperature for this layer is above room temperature. The ferromagnetic behavior is further confirmed when measuring the isotherm magnetization, where a clear hysteretic behavior is observed, see Fig. 3.

Fig. 3 (left and right) show the magnetization versus field reversal (M-H) measured at 5 and 300K, respectively, for all the samples. Hysteretic behavior is only observed in the Mn-RTA sample at both temperatures. The saturation moment ($M_s$) and the coercive field ($H_c$) for this sample are $7x10^{-4}$ emu/m$^{-3}$ and $2x10^{-2}$ Oe at 5 K and $5.1x10^{-4}$ emu/m$^{-3}$ and $1.5x10^{-2}$ Oe at 300 K, respectively.

A characterization of the structural properties of the samples has been done in order to investigate the changes after implantation and annealing as well as, the possible appearance of secondary phases and see the relation between structure and ferromagnetism in these samples, if any. X-ray diffraction patterns for as-grown, Mn and Ar-implanted and post-annealed layers are shown in Fig. 4. The four Bragg peaks observed in all XRD patterns indicate that films are composed of an InAs layer oriented along the (100) direction grown on a (100) GaAs substrate. An additional small peak located at 2θ= 45.09 º is observed for the Mn-RTA layer, which indicates the appearance of a secondary phase. This secondary phase is neither a Mn-As, Mn-As-In nor In-Mn combination, since according with the JCPDS database none of these phases present diffraction maxima around d=2.01Å.

The inset in Fig. 4 shows small changes in width and 2θ position of the (200) InAs peak for implanted and post-annealed layers in comparison to the reference film. These changes are related to implantation-induced damage and damage recovery after RTA. As shown in the inset in Fig. 4 the (200) InAs peak for the Mn- and Ar-implanted layer became broader than in the as-grown sample indicating a higher lattice disorder in implanted layers. After annealing the peak width slightly decreases in comparison to that of implanted layers, but it is still broader than the reference one, meaning that annealing under these conditions recovers only partially the implantation-induced damage. No shift in the (200) InAs peak position is detected for the Mn-implanted layer, which indicates that Mn implantation does not induce any additional stress in the InAs layer. This suggest, as next corroborated by PIXE measurement, that Mn occupies substitutional positions into the InAs lattice, whereas the peak shifts to higher angles for the Mn-RTA. A completely different situation is observed for the Ar-implanted and Ar-RTA layers, where a shift to lower angles in the (200) InAs peak position is observed.

In order to determine the location of Mn atoms in the InAs lattice in the Mn-implanted and Mn-RTA samples, RBS and PIXE measurements were performed in channeling configuration along the <100> and <110> axis. These measurements also allow us to check for the damage induced in the InAs lattice by implantation and damage recovery after annealing.[18] Because of the similarity in mass numbers of Mn and Ga, RBS data do not allow to resolve the Mn and Ga signals, which can be done well by PIXE. Therefore, PIXE was used for the determination of the Mn location in the film.

PIXE angular scan curves along <100> and <110> crystal axis taken from the In-L, Mn-$K_\alpha$+$K_\beta$ and As-$K_\beta$ for Mn-implanted and post annealed layers are shown in Fig. 5. For comparison an angular scan curve corresponding to a reference as-grown sample is also shown. The degree of lattice order of the as-grown layer is inferred from the measured minimum yield ($\chi_{min}$). In the as-grown layer the minimum yield for In at the surface is significantly higher than that expected for a typical single crystal, $\chi_{min}$ = 0.02-0.03,[18] which indicates that the surface of the 250 nm InAs layer is still affected by the typical defects such as misfit dislocations at the heterojunction, being not fully free of stress or defects.[19,20]

The large differences observed in $\chi_{min}$ from In and from As atoms along the <100> and <110> axis are due to the fact that the As signal obtained by PIXE comes from the As atoms of the InAs layer and also from the GaAs substrate. Indeed, Rutherford backscattering Spectroscopy measurements carried out in channeling configuration (RBS/C) (not shown) which offer depth resolution, reveal only small differences in $\chi_{min}$ for the In and As atoms located in the InAs layer. For this reason, and in order to obtain information only about the InAs layer, we will consider for the analysis of PIXE data only the In and Mn signals.

Following room temperature implantation the scan curves for Mn and In atoms along the <100> and <110> axis almost overlap, indicating that most of the Mn-implanted atoms occupy subsbtitutional positions into the In-site. The quantification of the percentage of In atoms displaced from their lattice position after implantation is estimated from the $\chi_{min}$ value. As observed in Fig. 5 $\chi_{min}$ for In atoms increases up to 0.52 for <100> and 0.64 for <110>, in comparison to 0.26 <100> and 0.48 <110> measured for the as-grown sample, which indicates that approximately 26% for the <100> and 18% for the <110> of the In atoms were displaced from their lattice position, being randomly distributed in the InAs lattice after implantation.

According to the angular scan curves, annealing drives the Mn atoms to move away from lattice positions to be randomly distributed in the InAs lattice. Moreover, the thermal treatment promotes damage improvement, reducing $\chi_{min}$ observed for In after implantation. As deduced from the changes in $\chi_{min}$ (values (0.36 for <100> and 0.35 for <110>) after annealing, damage recovery proves to be an anisotropic process. The fact that along the <110> axis $\chi_{min}$ for the annealed layer is smaller than that for the as-grown one would indicate that not only the radiation damage is diminished, but also part of lattice disorder induced by the as-grown defects, i.e. dislocations, whereas along the <100> axis only a small fraction of the radiation-induced damage (16%) is reduced. Thus, after annealing around 36% of the In atoms are still randomly distributed.

RBS data indicate that even sequential Mn implantation at fluences as high as $8x10^{16}$ $cm^{-2}$ produced high sputtering of the InAs layer, reducing up to a factor of 0.6 its thickness. Therefore, the density of Mn atoms that can be implanted into the InAs layer is limited by its low radiation-resistance.

In order to eliminate the possibility that the magnetism observed at room temperature is due to lattice disorder instead of the segregation of secondary phases of Mn, a set of samples were implanted with Ar, which is inert, at the same fluences and under conditions, which according to SRIM calculations[17] give rise to similar damage patterns as for Mn (similar densities of atomic displacement and vacancies creation. As shown in Fig. 2 and in Fig. 3 neither the Ar-implanted nor the Ar-RTA layers samples present any magnetic behavior, indicating that magnetism is not related to lattice disorder but to magnetic dopants, and in particular to the formation of a Mn-based second phase.

The depth distribution of the randomly distributed Mn atoms in the Mn-RTA layer can be inferred by comparing experimental and simulated RBS spectra. These data show that after the thermal treatment manganese is not uniformly distributed along the whole InAs layer, but it is segregated near to or at the film surface. Thus, Mn is supposed to be in direct contact with air and therefore, the combination of Mn with O should also be taken into account. Indeed, the elemental composition of this Mn-O surface layers is estimated from RBS data to be 34 $_{at.}$% Mn and 66 $_{at.}$% O which corresponds to a general formula of $MnO_{1.94}$. Taking in mind that a Mn-O phase could be formed, we check the X-ray database and one Mn-O phase present a diffraction maxima at $2\theta= 45.09^{o}$, the oxygen deficient $MnO_{1.937}$ phase,[21] which oxygen concentration is close to the obtained value by RBS. The segregated phase must be oriented along the (100) direction, the same as the substrate and the film, since is the (400) maximum is the only one observed and no the more intense peak of this phase, which would be the more probable one if the segregated $MnO_{1.94}$ nanoparticles were randomly oriented. This phase is the cubic defect spinel λ-$MnO_2$. This oxygen-deficient-$MnO_2$ polymorph with a lattice constant of 8.04 Å and a space group of Fd3m. Given that the only sample which exhibits a ferromagnetic behavior is the one containing this peak, the magnetic response might be due to this phase. Since the phase is oxygen deficient $Mn^{3+}$ and $Mn^{4+}$ should coexist in the sample. The coexistence of $Mn^{3+}$-O-$Mn^{4+}$ bonds is known to be responsible for FM via the double-exchange mechanism in different compounds[3,22,23], which usually exhibits a large Curie temperature ($T_C$) even above room temperature.[24] This behavior is similar to what has been previously observed for the ZnO/$MnO_2$ system.[3] Other possible explanation of the magnetism observed can be consistent with the geometric frustration inherent in the Mn sublattice in this phase.[25,26] These experimental data contrast somewhat with those described in literature in which the ferromagnetic behavior at room temperature of $In_{1-x}Mn_xAs$ layers has been traditionally associated with the formation of magnetic MnAs-clusters[10] or to the doping introduction of Mn inside the InAs structure.

## CONCLUSIONS

The magnetic properties of Mn and Ar- implanted and post-annealed (Mn-RTA and Ar-RTA) InAs layers grown by MBE has been investigated. Mn-InAs samples where Mn is substitutional in the In-site behave diamagnetically. Room temperature ferromagnetic-like behavior was only present in the Mn-RTA layer. The magnetic behavior of this layer (in which the Mn atoms are not substitutional inside the InAs lattice and mainly located near to or at the InAs surface layer) could be related to the formation of a second phase which has been identified from RBS and XRD to be an oxygen deficient $\lambda$-$MnO_{1.94}$ phase. Changes in the width and position of Bragg diffraction peaks and in the $\chi_{min}$ value in the PIXE angular scan curves evidence that implantation induced lattice disorder in the InAs lattice, which is partially or fully recovered after annealing. No significant stress is developed in Mn-implanted layers in which almost all Mn atoms occupy substitutional positions at the In site, while a tensile stress is observed for the Mn-RTA layer in which the Mn atoms are segregated. A compressive stress has been measured in the Ar-implanted samples as well as in the Ar-RTA layer. The density of the Mn atoms which can be introduced via implantation in the InAs layer is limited by its low radiation-resistance. By comparing magnetic and structural data for Mn and Ar implanted layers, no relation between magnetism and lattice disorder has been detected.

## ACKNOWLEDGMENTS

This work was supported by CSIC 2006-50F0122, CSIC 2007-50I015, and in part by the MAT2005-06024-C02-01 project. R. G. A acknowledges to the MEC and CSIC for the Juan de la Cierva financial support. The PIXE measurements have been supported by the EU-“Research Infrastructures Transnational Access” program AIM “Center for Application of Ion Beams in Materials Research” under EC contract no. 025646.

## Figure captions

**Fig. 1**.- Mn (Left) and Ar (right) depth distribution within the InAs layer after implantation as simulated by the TRIM Montecarlo code[17].

**Fig. 2**.- Temperature dependence of the magnetization at constant field for the reference (black stairs), Mn-implanted (black circles), Mn-RTA (open circles), Ar-implanted (black squares), and Ar-RTA (open squares) samples.

**Fig. 3.-** Magnetization versus field reversal (M-H) measured at 5 K(right) and 300 K (left) for reference ( black stairs), Mn-implanted (black circles), Mn-RTA (open circles), Ar-implanted (black squares), and Ar-RTA (open squares) samples.

**Fig. 4.-** Logarithmic scale picture of the XRD patterns for the reference (black dash line), Mn-implanted (crosses), Mn-RTA (black dots), Ar-implanted (stairs), and Ar-RTA (transversal black lines) samples. The insets show a zoom of the (200) InAs peak for all the previously described layers. The black line placed in the insets at $2\theta = 29.455$ º corresponds to the diffraction angle for an ideal Bulk InAs sample.

**Fig. 5.-** PIXE angular scan curves along the <100> (right) and <110> (left) axis of the As-$K_\beta$ (crosses), In-L (black circles) and Mn-$K_\alpha$+$K_\beta$ (open circles) lines for the reference, as-grown, (top graphs), Mn-implanted (middle graphs) and Mn-RTA (bottom graphs) layers.

## REFERENCES

[1] G. A. Prinz, Science, **282**, 1660 (1998)

[2] T. Dietl, H. Ohno, F. Matsukura, J. Cibert and D. Ferrand. *Science* **287,** 101 (2000)

[3] M. A. García, M. L. Ruiz-González, A. Quesada, J. L. Costa-Krämer, J. F., Fernández, S. J. Khatib, A. Wennberg, A. C. Caballero, M. S. Martín-González, M. Villegas, F. Briones, J. M. González-Calbet, and A. Hernando, Phys. Rev. Lett. **94**, 217206 (2005)

[4] A. Quesada, M. A. Garcia, M. Andres, A. Hernando, J. F. Fernandez, A. C. Caballero, M. S. Martin-Gonzalez, and F. Briones, J. Appl. Phys. **100**, 113909 (2006)

[5] J. M. D. Coey, Curr. Opin. Solid State Mater. Sci. **10**, 83 (2006).

[6] M. S. Martín-González, J. F. Fernández, F. Rubio-Marcos, I. Lorite, J. L. Costa-Krämer, A. Quesada, M. A. Bañares, and J. L. G. Fierro J. Appl. Phys. **103**, 083905 (2008)

[7] A. J. Blattner and B. W. Wessels, Appl. Surf. Sci. **221**, 155 (2004)

[8] P. T. Chiu, B. W. Wessels, D. J. Keavney and J. W. Freeland Appl. Phys. Lett. **86**, 072505 (2005)

[9] H. Ohno, Science **281**, 951 (1998)

[10] H. Munekata, H. Ohno, S. von Molnar, Armin Segmüller, L. L. Chang, and L. Esaki, Phys. Rev. Lett. **63**, 1849 (1989)

[11] P. J. Wellmann, W. V. Schoenfeld, J. M. Garcia, and P. M. Petroff, J. of Electronic Materials, **27**, 1030 (1986)

[12] J. Shi, J. M. Kikkwa, D. D. Awschalom, G. Medeiros-Ribeiro, P. M. Petroff, and K. Babcock, J. Appl. Phys. **79**, 5296 (1996)

[13] N. Theodoropulo, F. Hebard, S. N. G. Chu, M. E.Overberg, C. R. Abernathy, S. J. Pearton, R. G. Wilson, J. M. Zavada, J. Appl. Phys. **91**, 7 (2002)

[14] H. Ofuchi, T. Kubo, M. Tabuchi, Y. Takeda, F. Matsukura, S. P. Guo, A. Shen, and H. Ohno, J. Appl. Phys. **89**, 66 (2001)

[15] A. Krol, Y. L. Soo, S. Huang, Z. H. Ming, Y. H. Kao, H. Munekata, and L. L. Chang, Phys. Rev. B. **47**, 7187 (1993)

[16] F. Briones, L. González, and A. Ruiz, Appl. Phys. A **49,** (1989) 729.

[17] J. F. Ziegler, M.D. Ziegler, J.P.Biersack SRIM-2006.02

[18] J. R. Tesmer, M. Nastasi, handbook of Modern Ion Beam Materials Analysis, MRS (1995)

[19] C. Chang, C. M. Serrano, L. C. Chang, L. Esaki, Appl. Phys. Lett. **37**, 538 (1980)

[20] R. S. Williams, B. M. Paine, W. J. Schaffer, S. P. Kowalczyk, Jour. Vac. Sci. Technol. **21**, 386 (1982)

[21] JCPDS 421169

[22] M. B. Salamon and M. Jaime, Rev. Mod. Phys. **73**, 583 (2001)

[23] J. Alonso, A. Arroyo, J.M. González-Calbet, M. Vallet-Regi, J.L. Martínez, J.M. Rojo, and A. Hernando, Phys. Rev. B **64**, 172410 (2001).

[24] T. Okuda, A. Asamitsu, Y. Tomioka, T. Kimura, Y. Taguchi, and Y. Tokura Phys. Rev. Lett. **81**, 3203 (1998).

[25] J.E. Greedan, N.P. Raju, A.S. Wills, C. Morin, and S.M. Shaw Chem. Mater. **10**, 3058 (1998)

[26] N. Wang, X. Cao, G. Lin, and Y. Shihe Nanotech. 18 475605 (2007)

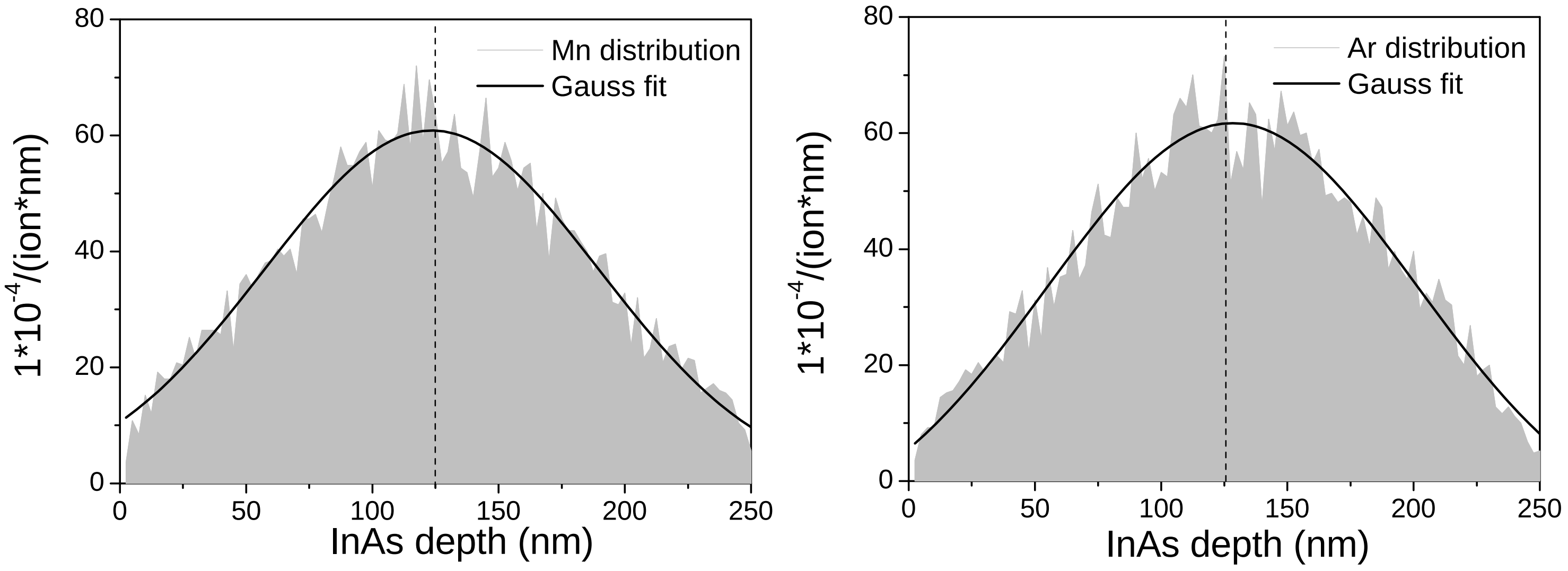

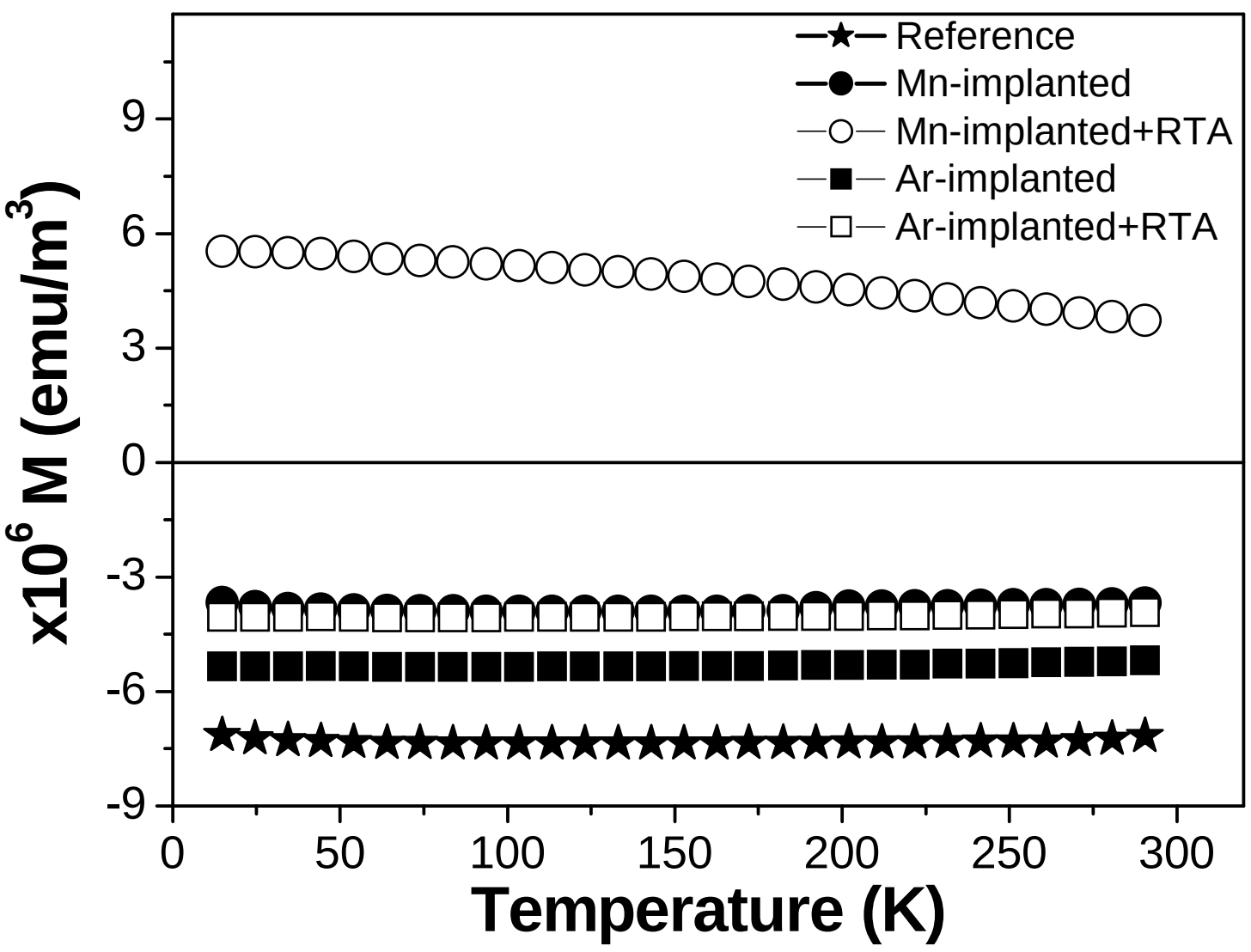

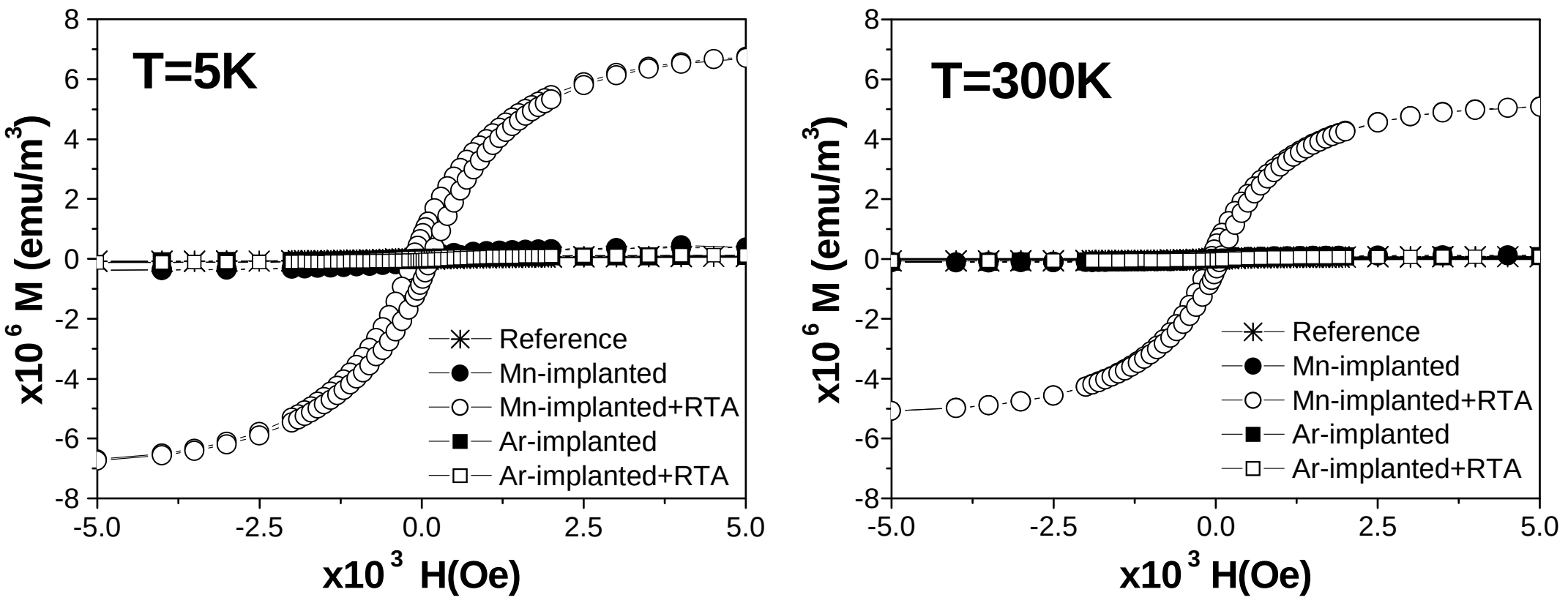

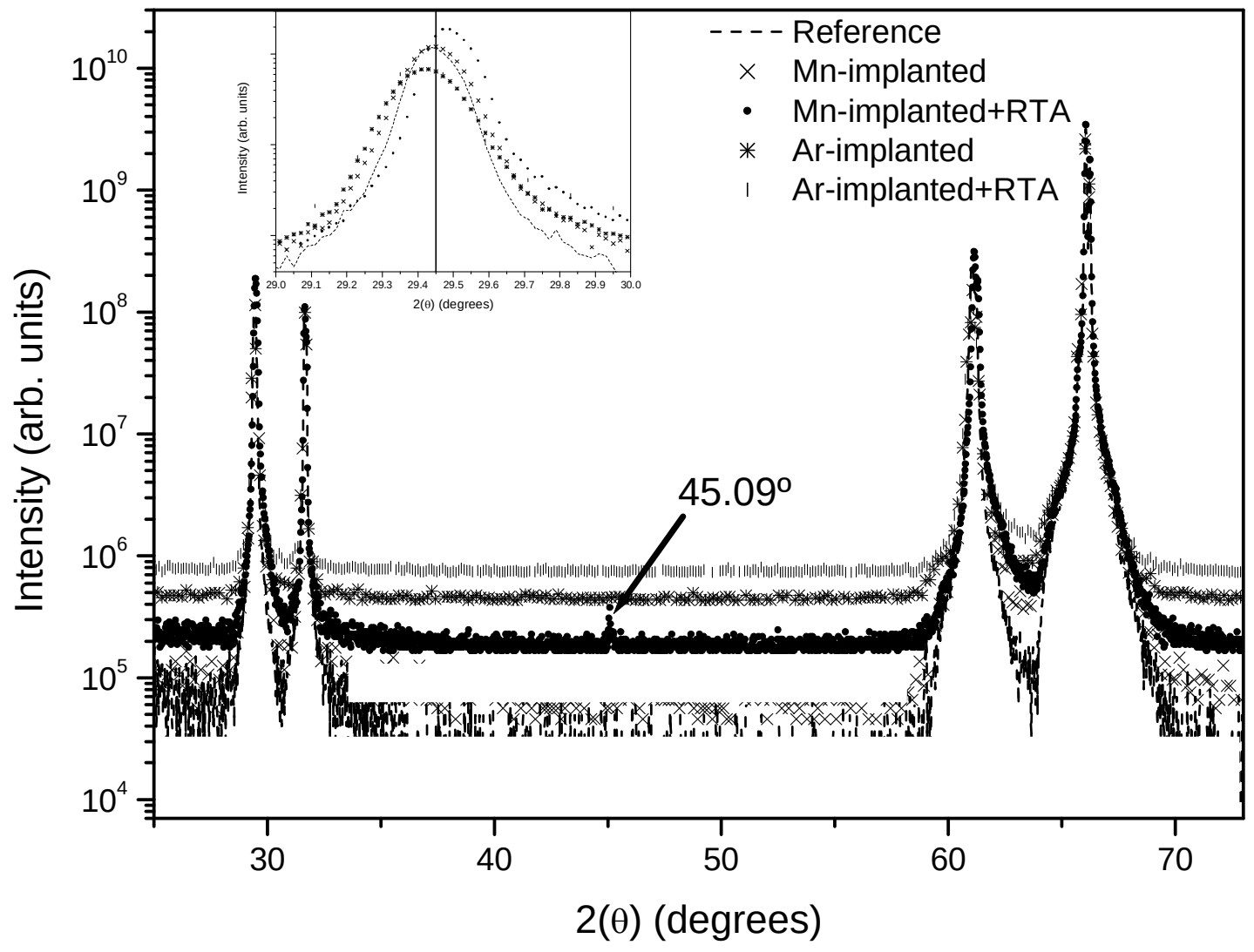

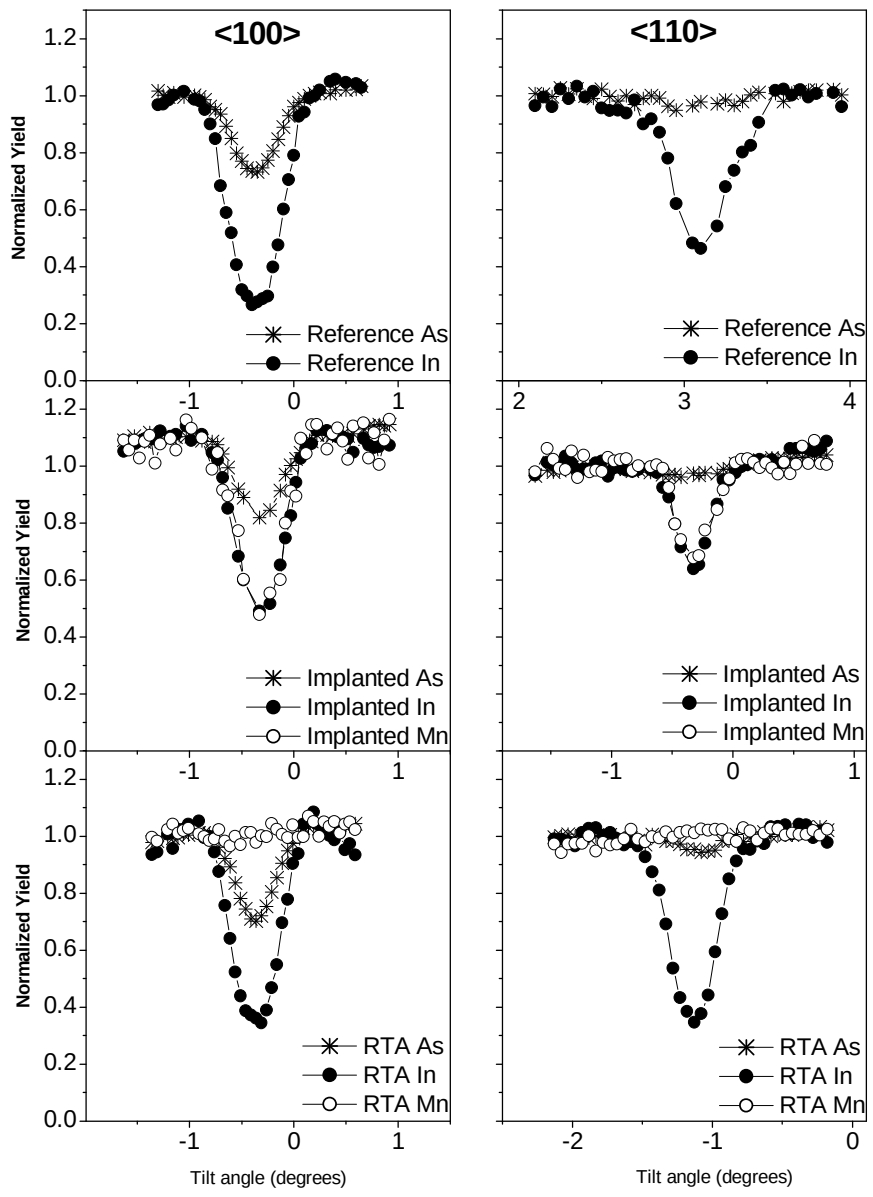